\documentclass[aip,twocolumn,showpacs,superscriptaddress,reprint,amsmath,amssymb,floatfix,nofootinbib]{revtex4-1}
\usepackage{graphicx}
\usepackage{bm}
\usepackage{amssymb}
\usepackage{epsfig}
\usepackage{pdfsync}

\begin{document}
\title{Geometric reduction of dynamical nonlocality in nanoscale quantum circuits}

\author{E.~Strambini$^*$}
\email{e.strambini@sns.it}
\affiliation{NanoElectronics Group, MESA+ Institute for Nanotechnology, University of Twente, P.O. Box 217, 7500 AE, Enschede, The Netherlands}
\author{K.~S.~Makarenko$^*$}
\affiliation{NanoElectronics Group, MESA+ Institute for Nanotechnology, University of Twente, P.O. Box 217, 7500 AE, Enschede, The Netherlands}
\author{G.~Abulizi}
\affiliation{NanoElectronics Group, MESA+ Institute for Nanotechnology, University of Twente, P.O. Box 217, 7500 AE, Enschede, The Netherlands}
\author{M.P.~de~Jong}
\affiliation{NanoElectronics Group, MESA+ Institute for Nanotechnology, University of Twente, P.O. Box 217, 7500 AE, Enschede, The Netherlands}
\author{W.~G.~van~der~Wiel}
\email{W.G.vanderWiel@utwente.nl}
\affiliation{NanoElectronics Group, MESA+ Institute for Nanotechnology, University of Twente, P.O. Box 217, 7500 AE, Enschede, The Netherlands}
\let\thefootnote\relax\footnote{* These authors contributed equally to this work.}
\begin{abstract}
Nonlocality is a key feature discriminating quantum and classical physics.
Quantum-interference phenomena, such as Young's double slit experiment, are one of the clearest manifestations of nonlocality, recently addressed as \emph{dynamical} to specify its origin in the quantum equations of motion.
It is well known that loss of dynamical nonlocality can occur due to (partial) collapse of the wavefunction due to a measurement, such as which-path detection. 
However, alternative mechanisms affecting dynamical nonlocality have hardly been considered, although of crucial importance in many schemes for quantum information processing.
Here, we present a fundamentally different pathway of losing dynamical nonlocality, demonstrating that the detailed geometry of the detection scheme is crucial to preserve nonlocality. 
By means of a solid-state quantum-interference experiment we quantify this effect in a diffusive system. 
We show that interference is not only affected by decoherence, but also by a loss of dynamical nonlocality based on a local reduction of the number of quantum conduction channels of the interferometer. 
With our measurements and theoretical model we demonstrate that this mechanism is an intrinsic property of quantum dynamics.
Understanding the geometrical constraints protecting nonlocality is crucial when designing quantum networks for quantum information processing.

\end{abstract}
\maketitle
Besides the famous Bell nonlocality~\cite{bell_einstein-podolsky-rosen_1964}, allowing for correlations between distant particles stronger than classically possible, the importance of another fundamental quantum nonlocality, although discovered over five decades ago, has been pointed out recently, and has been referred to as ``dynamical nonlocality''~\cite{Popescu_Dynamical_2010, Popescu_Nonlocality_2014}. 
This dynamical nonlocality follows from the nonlocality of the quantum equations of motion, clearly manifests itself in branching geometries, and is at the base of all quantum-interference phenomena, such as Young's double slit experiments, Mach$-$Zehnder interferometers~\cite{Klepp_Fundamental_2014}, and the Aharonov-Bohm (AB) effect~\cite{Aharonov_Significance_1959}.
Loss of dynamical nonlocality implies quenching of interference.
If a quantum mechanical wavefunction, e.g. describing a propagating electron, is spatially split to enclose a region of magnetic flux, a phase difference develops, leading to quantum-interference oscillations periodic in the magnetic flux quantum, $h/e$ ($h$ is Planck's constant and $e$ is the electronic charge).
The dynamical nonlocality of the wavefunction induced by a branching geometry is, together with quantum coherence, the basic ingredient of any quantum-interference phenomenon.

\begin{figure*}[!ht]
\begin{center}
\includegraphics[width=0.85\textwidth,clip=true]{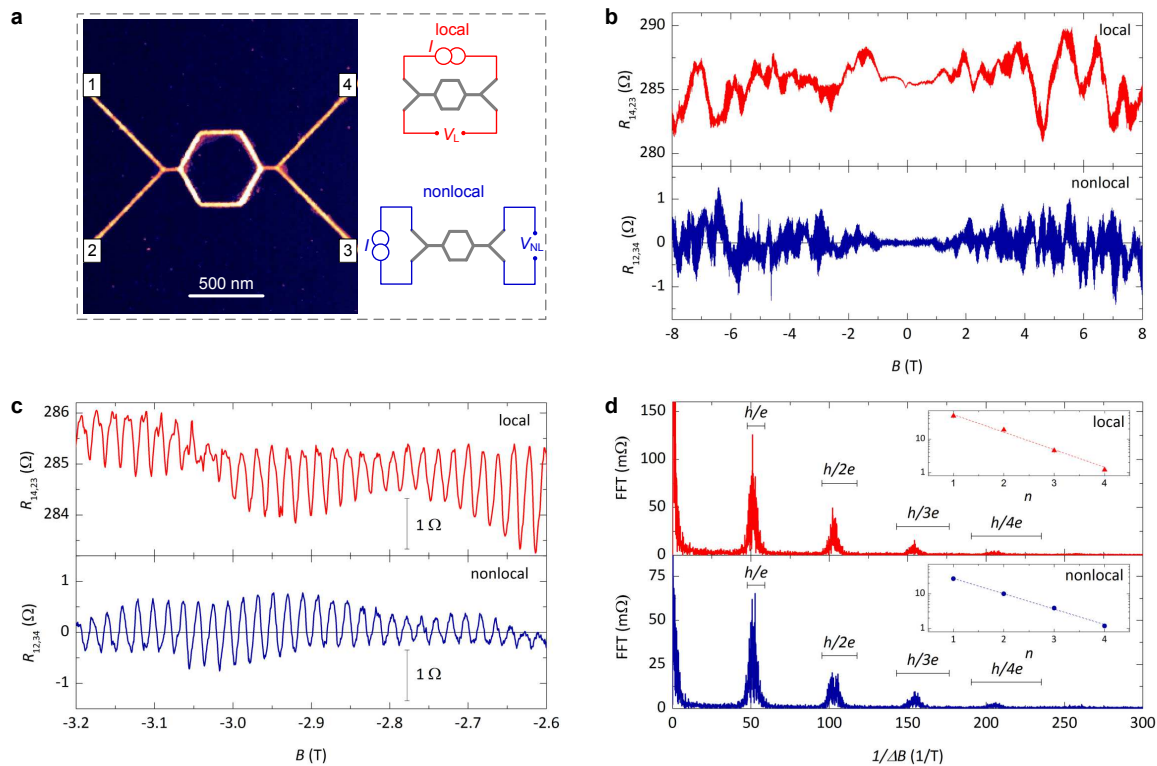}
\caption{\label{GoldenRing} \textbf{Comparison between local and nonlocal AB interference. a}, Atomic force microscopy (AFM) image of the device and schematics of two measurement configurations: local 4-terminal ($R_{14,23}$, current: from 1 to 4, voltage: between 2 and 3) and nonlocal 4-terminal ($R_{12,34}$, current: from 1 to 2, voltage: between 3 and 4). 
\textbf{b}, Magnetoresistance of the ring in \textbf{a}, measured in local and nonlocal setups at 40~mK and with 30~nA ac excitation current. 
\textbf{c}, Details of the magnetoresistances represented in \textbf{b}. 
\textbf{d}, Fourier transform of the data in \textbf{b}. The bounds for the flux periods \textit{h/ne} ($n$ = 1, 2, 3, 4) are calculated based on the inner and outer circumference of the device. The insets show the dependence of the peak amplitude in $n$ (dots) with the respective exponential fit (dash line).}
\end{center}
\end{figure*}

In this Report, we present a set of electron-interference experiments in which, besides the effect of decoherence, i.e. the randomization of the single-electron phase due to inelastic scattering, we also observe and quantify the effect of loss of dynamical nonlocality~\cite{Popescu_Dynamical_2010, Popescu_Nonlocality_2014}.
Loss of dynamical nonlocality has been first demonstrated and tuned in which-path experiments~\cite{Zou_Induced_1991, Buks_Dephasing_1998}, one of the most remarkable manifestations of the complementarity principle. In these experiments, detection of the photon, respectively electron, traversing the two branches of the interferometer induces a collapse of the delocalized quantum state (wave-like) in a localized one (particle-like), resulting in quenching of the interference.
Here we investigate a new phenomenon affecting the dynamical nonlocality of the electron that, different from the which-path scheme, is inherent to the device geometry and originates from a reduction of the number of conduction channels instead of from the interaction with the external environment described above.
This loss of dynamical nonlocality is not observed in a regular, local (L) AB measurement geometry (see Fig.~\ref{GoldenRing}a(red)), but does play a significant role in the nonlocal (NL) geometry of Fig.~\ref{GoldenRing}a(blue). In the nonlocal geometry, also the quantum nature of the \emph{side arms} of the ring is crucial, as opposed to the local geometry, where only the ring itself matters. In the side arms, conduction channels linked to the two different branches of the ring can directly interact and merge, which is prevented in the ring itself. The implication of this fundamental difference will be discussed in detail below.

Figure~\ref{GoldenRing}a shows an image of a representative Au AB interferometer. More than ten devices of different size exhibited similar behaviour (more results in the Supplementary Information). The AB ring is connected to two $\sim$100~nm long side arms, which each split into two leads. This device layout allows for measuring the differential resistance ($dV_\mathrm{L}/dI \equiv R_\mathrm{L} = R_{14,23}$) in the conventional, local configuration (Fig$.$~\ref{GoldenRing}a(red)). The local signal shown in the upper panel of Fig$.$~\ref{GoldenRing}b shows clear AB oscillations, see Fig$.$ 1c, superimposed on a reproducible background of universal conductance fluctuations (UCF)~\cite{benoit_length-independent_1987}. The corresponding Fourier spectrum (Fig$.$~\ref{GoldenRing}d) clearly reveals the AB period $h/e$, as well as higher harmonics up to $h/4e$. 

In addition to $R_\mathrm{L}$, we have also measured the \emph{nonlocal}  differential resistance $dV_\mathrm{NL}/dI \equiv R_\mathrm{NL} = R_{12,34}$ according to the scheme in Fig$.$~\ref{GoldenRing}a(blue). Classically, one expects to measure a zero signal in this configuration (or exponentially small according to the Van der Pauw theorem~\cite{van_der_pauw_method_1958}). However, in the quantum coherent transport regime, the signal may be nonzero~\cite{benoit_length-independent_1987, DiVincenzo_Voltage_1988} as the electron wave function, dynamically delocalized in the system, can connect the current and voltage probes within a distance of a few coherence lengths.
Indeed a nonlocal AB effect has been measured in \emph{ballistic} rings~\cite{Kobayashi_Probe-configuration-dependent_2002, Buchholz_Nonlocal_2009, Buchholz_Control_2010,Lin_Temperature-_2010,Lin_Asymmetric_2011}, while for diffusive metals the presence of an Au ring side-connected to a nanowire was detected in the oscillations of the local resistance~\cite{Umbach_Nonlocal_1987}.
To the best of our knowledge, nonlocal AB oscillations have never been observed in diffusive metallic rings and are often neglected in the theoretical estimations~\cite{Buttiker_Four-Terminal_1986}. In our diffusive ring, with the design shown in Fig$.$~\ref{GoldenRing}a, we do however observe very clear nonlocal AB oscillations (Fig$.$~\ref{GoldenRing}c) around a zero average (lower panel of Fig$.$~\ref{GoldenRing}b). The Fourier spectrum shows the same four harmonics as for the local signal, but with roughly half the amplitude (Fig$.$~\ref{GoldenRing}d, lower panel).

\begin{figure*}[!ht]
\begin{center}
\includegraphics[width=0.85\textwidth,clip=true]{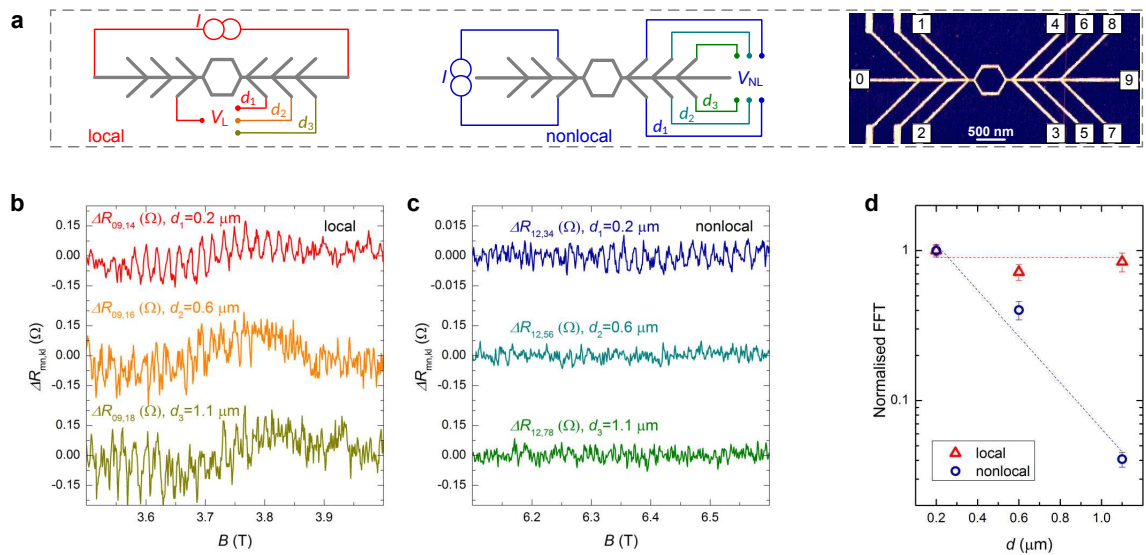}
\caption{\label{Fishbone} \textbf{Length dependence of the local and nonlocal AB interference. 
a}, schematics of two 4-terminal configurations and AFM image of the multi-terminal device.
\textbf{b} and \textbf{c}, Comparison between the oscillations of the magnetoresistance measured at different distances $d$ (where $d$ is the total length of both side arms) in the local and nonlocal configuration.
\textbf{d} Amplitude of the local (triangles) and nonlocal (circles) AB oscillations as a function of $d$. For better comparison, the data are normalized by the amplitude at $d = d_1$.}
\end{center}
\end{figure*}

For both the local and nonlocal configurations the amplitude of these Fourier peaks follows a clear exponential decay in the harmonic index $n$ (as shown in the insets of Fig$.$~\ref{GoldenRing}d), as the path length scales with $n$, corresponding to the number of times that the electron encircles the ring. This decay thus provides a direct measure of the electron coherence length $l_{\varphi}$~\cite{Hansen_Mesoscopic_2001}.  Using the power law $e^{ -0.6 \frac{n C}{l_{\varphi}} }$, expected for a diffusive interferometer~\cite{DiVincenzo_Voltage_1988}, where $C$ is the ring circumference, we find $ l_{\varphi} \simeq$1.5~$\mu$m for both the local and nonlocal configuration. The value of $l_{\varphi}$ is consistent with  values found  from weak localization and UCF measurements in mesoscopic Au structures at millikelvin temperatures~\cite{Saminadayar_Electron_2007}. Other similar rings (Supplementary Information) show the same trend with $l_{\varphi}$ = 1-2~$\mu$m, independent of sample size and temperature (below 1~K), in agreement with other studies for gold~\cite{mohanty_intrinsic_1997}.

Our results demonstrate that both the local and nonlocal interference are equally suppressed by decoherence inside the ring. Decoherence inside the side arms is only relevant for the nonlocal signal, leading to an additional damping $e^{-1.1 \frac{d}{l_{\varphi}}}$, where $d$ is the total length of the two side arms together~\cite{DiVincenzo_Voltage_1988}. As a consequence, the amplitude of the nonlocal AB oscillations is smaller than the local AB oscillations, as seen in Fig$.$~\ref{GoldenRing}. This is in agreement with other studies on nonlocal interference~\cite{benoit_length-independent_1987, Haucke_Universal_1990, Kobayashi_Probe-configuration-dependent_2002}. 
To study the damping effect of the side arms on the nonlocal AB oscillations in more detail, we measured $R_\mathrm{NL}$ and $R_\mathrm{L}$ in the multi-terminal interferometer shown in Fig$.$~\ref{Fishbone}a, in which $d$ can be varied.

Clear AB oscillations are visible for $d$ = 200~nm ($d_1$) in both $R_\mathrm{L}$ ($R_{09,14}$, red plot in Fig$.$~\ref{Fishbone}b) and $R_\mathrm{NL}$ ($R_{12,34}$, blue plot in Fig$.$~\ref{Fishbone}c). For $R_\mathrm{L}$ the amplitude of the AB oscillations does not significantly change when the side-arms length $d$ increases, see Fig$.$~\ref{Fishbone}d. By contrast, the amplitude of the nonlocal AB oscillations is strongly suppressed with increasing $d$. The damping of the nonlocal oscillations can be described by the power low $e^{-\frac{d}{\xi}}$, see Fig$.$~\ref{Fishbone}d, where $\xi \sim 350 \pm 50$~nm is the characteristic length scale of the damping. Interestingly, this damping is about three times faster than the theoretical expectation for a nonlocal measurement ($e^{-\frac{1.1 d}{l_{\varphi}}}$)~\cite{DiVincenzo_Voltage_1988}, which would give a value of $1.4 \pm 0.2 \mu$m, but still consistent with the damping measured in the UCF of Au nanowires~\cite{Haucke_Universal_1990}.
\begin{figure}[!ht]
\includegraphics[width=\columnwidth,clip=true]{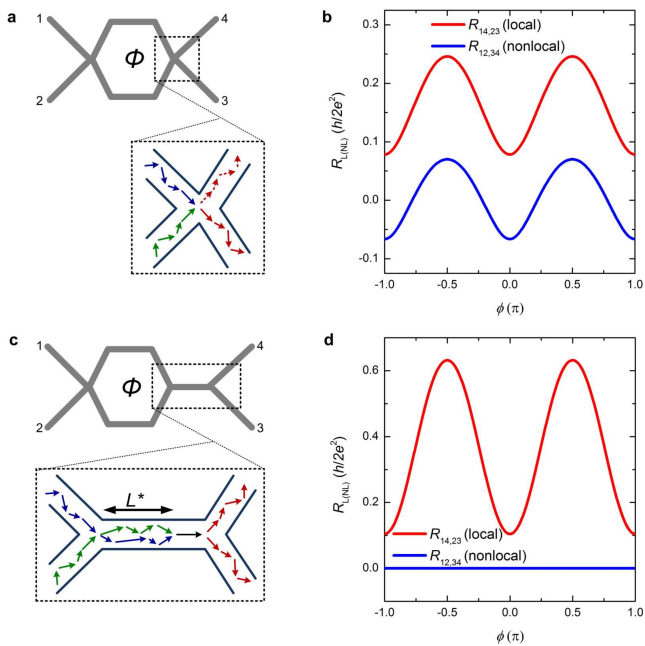}
\caption{\label{Model} \textbf{Consequence of the reduction of the amount of quantum conduction channels in the side arm. a} Scheme of the simple 4-terminal AB interferometer with a detailed representation of the second node of the interferometer in which the two electronic paths of the ring (green and blue arrows) interfere generating the two phase-correlated output channels represented by the red arrows (solid and dashed).
\textbf{b}, Comparison between the 4-terminal local (red) and nonlocal (blue) resistances simulated as a function of the magnetic AB phase $\phi$ for an AB ring without side arms. 
\textbf{c}, Scheme of the 4-terminal AB interferometer with one single-channel side arm allowing a reduction of two conduction channels into one after the characteristic distance $L^*$ and leading to quenching of the phase correlations between the two electronic paths of the ring, resulting in two phase-uncorrelated output channels represented by the solid red arrows.
\textbf{d}, Comparison between the 4-terminal local (red) and nonlocal (blue) resistances simulated as a function of the magnetic AB phase $\phi$ for the AB ring with one single-channel side arm.}
\end{figure}

We ascribe this large discrepancy to a loss of dynamical delocalization in the nonlocal geometry. To catch the essential physics describing this loss we use the Landauer-B\"{u}ttiker (LB) formalism~\cite{Buttiker_Four-Terminal_1986} to model the magnetoresistance of the device first in the few-channel regime, and then we extend the argument to our multi-channel diffusive experimental system. 
The magnetoresistance of the ring, represented in Fig$.$~\ref{Model}a and modelled in the single-channel approximation (see Methods), well describes the fundamental characteristics of the experimentally observed $R_\mathrm{L}$ and $R_\mathrm{NL}$: a positive $R_\mathrm{L}$ exhibiting AB oscillations, and $R_\mathrm{NL}$ oscillating between positive and negative values with a similar amplitude ($\Delta R_\mathrm{NL} \simeq \Delta R_\mathrm{L}$), as shown in Fig$.$~\ref{Model}b.
This picture changes drastically when modelling the same ring with an additional side arm, as represented in Fig$.$~\ref{Model}c. The addition of the one-dimensional, i.e. single-channel side arm, results in the complete quenching of all $R_\mathrm{NL}$, while the behaviour of $R_\mathrm{L}$ is only weakly affected, see Fig$.$~\ref{Model}d. 
We propose that this geometrical localization, imposed by the side arm, is essentially the origin of the suppression observed in the nonlocal interference and reduces when the side arm is composed of more than one channel, as is expected in the diffusive case.

This hypothesis is also in good agreement with experimental observations of the nonlocal AB effect in ballistic systems~\cite{Kobayashi_Probe-configuration-dependent_2002,Lin_Temperature-_2010,Lin_Asymmetric_2011,Buchholz_Nonlocal_2009, Buchholz_Control_2010}, in which the nonlocal AB oscillations have been reported only for several transport channels.
In our metallic devices, the side arms are composed of a multitude of interacting channels due to the diffusive electron motion, and the probability that scattering merges two otherwise distinct conduction channels into a single one (event represented in the scheme of Fig$.$\ref{Model}c) is high. The measured signal results from the average of all the possible diffusive electron paths. In this average the paths that experience such a reduction do not contribute to interference, as demonstrated in our model. Due to the random distribution of electron paths in a diffusive metal, the probability for this event to occur is expected to scale exponentially with the length of the side arm ($\propto 1-e^{-d/L^*}$, where $L^*$ is the characteristic length scale of this event). As a consequence the nonlocal signal is damped within the same length scale ($e^{-d/L^*}$), additionally to the conventional damping induced by decoherence ($e^{\frac{1.1}{l_\varphi}}$)\cite{DiVincenzo_Voltage_1988}. 
This behaviour is in agreement with the strong exponential damping of the $R_\mathrm{NL}$ shown in Fig$.$~\ref{Fishbone}d from which we can estimate the characteristic length $L^* = (\frac{1}{\xi} - \frac{1.1}{l_\varphi})^{-1} \simeq 450\pm 50 $~nm.

We have thus shown a new aspect of the \emph{dynamical} nonlocality of electrons in a quantum nanoscale circuit that is solely governed by geometric aspects and not by external measurement. 
Understanding this geometrical constraint is essential for the optimal design of any quantum circuit based on the dynamical nonlocality of the electron.
Moreover, as deduced from our model, the specific link found between geometry and nonlocality is not limited by the diffusive transport regime of our experiment, nor by the specific quantum wave-particle used (in our case electrons) but, as other important geometrical constrains~\cite{Klepp_Fundamental_2014}, is a universal property of quantum dynamics. We believe that these results will trigger further investigation of the fundamental properties of quantum dynamics and of its application in nanoscale quantum circuits.

\section{Methods}
\subsection{Device fabrication.}
A 35-nm-thick SiO$_2$ layer was thermally grown on 4-inch p$^{++}$ Si wafers in order to electrically insulate the devices from the substrate. Afterwards the wafers were diced to a size of 10x10 mm$^2$ to fit the dilution refrigerator chip holder. The interferometer structures were patterned with electron-beam lithography (EBL), each sample was coated with 80 nm poly(methylmethacrylate) (PMMA-A2) resist and using a RAITH150-TWO system at an acceleration voltage of 20 kV and an aperture of 10 $\mu$m. EBL patterning was followed by electron-beam evaporation of a Ti (3 nm)/Au (30 nm) metal double layer and lift-off. Non-destructive imaging was performed with atomic force microscopy. 
\subsection{Experimental setup.}
The magnetoresistance of the AB interferometer, the temperature and excitation-current dependences were performed using a cryogen free dilution refrigerator (Oxford Instruments TritonTM 200) with a base temperature of 34~mK. A closed-cycle 3He cryostat with a base temperature of 250~mK (Oxford Instruments Heliox VL) was used during the measurements of the dependence on the probe configurations. In order to reduce 1/$f$ noise and interference a lock-in amplifier (Stanford Research Systems SR830) was used at a frequency of 17.77 Hz together with a low-noise current source and voltage amplifier (modules S4c and M2b, Quantum Transport, TU Delft, designed by Ing. Raymond Schouten). A custom-made LabView program was used to collect and analyse the data.
\subsection{Computational details}
In the LB formalism the transport properties of a 4-terminal device are completely described by a unitary $4\times4$ scattering matrix $t_{ij}$, representing the amplitude probabilities for an electron injected in the terminal $i$ to be transmitted to terminal $j$.
According to the scheme in Fig$.$~\ref{Model}a, the scattering matrix of a 4-terminal AB interferometer (without side arms) can be obtained by combining the scattering matrices of two $4\times4$ side nodes with a $4\times4$ phase-shifter representing the two branches of the ring, each of them shifting the electron phase by $\pm \phi /2$, where $\phi$ is the magnetic AB phase (see the Supplementary information for more details).
The 4-terminal resistances $R_{nm,kl}$ (current: from $n$ to $m$, voltage: between $k$ and $l$) are then extracted from the coefficients of the probability matrix $T_{ij} = |t_{ij}|^2$ according to the LB analytical formula~\cite{Buttiker_Four-Terminal_1986} 
\begin{equation}
\label{4R}
R_{nm,kl}= \frac{h}{2e^2} \frac{(T_{km} T_{ln}-T_{kn} T_{lm} )}{ \mathcal{D}}, 
\end{equation}
where $\mathcal{D}$ is a quantity including all of the $T_{ij}$ coefficients. Assuming a symmetric splitting probability in the two branches of the ring and an exponential damping of the amplitude ($e^{- \frac{C}{l_{\varphi}}}$) to simulate decoherence, we obtain the $R_\mathrm{L}(\phi$) and $R_\mathrm{NL}(\phi)$ plots shown in Fig$.$~\ref{Model}b.

With the same formalism we obtain the scattering matrix of the 4-terminal AB interferometer with a the single-channel side arm, schemed in Fig$.$~\ref{Model}c.
This forced reduction of conduction channels when the branches of the ring come together in the side arm generates an intrinsic constraint among the transmission coefficients representing the electrons injected in the device (e$.$g$.$ $t_{13} , t_{14}, t_{23} ,t_{24}$ in the schematic of Fig$.$~\ref{Model}c). As derived in the Supplementary Information, for this geometry the simple relation $t_{13}t_{24} = t_{23}t_{14}$ holds, independent of the complexity or asymmetry of the system. From the definition of the 4-terminal resistance (Eq$.$~\ref{4R}), it is straightforward to note that this geometrical constraint is the origin of the quenching of all the nonlocal resistances (e.g. $R_\mathrm{NL}= R_{43,12} \propto T_{13}T_{24}-T_{14}T_{23}=0$), as shown in Fi$.$~\ref{Model}d, while it is only weakly affecting the local one.
\section{Acknowledgements}
The authors thank Fabio Taddei, Vittorio Giovannetti, Vincenzo Piazza, Floris Zwanenburg and Kurt Vergeer for fruitful discussions. This work was financially supported by the European Research Council, ERC Starting Grant no. 240433 and through the EC FP7-ICT initiative under Project SiAM No 610637.
\section{Author contributions}
E.S. and K.S.M. carried out the experiments and performed the data analysis. E.S. developed theoretical model and performed simulations. K.S.M. fabricated samples. W.G.v.d.W. conceived the experiments, planned and supervised the project. M.P.d.J. contributed to the supervision. G.A. participated in carrying out the experiments. All authors discussed the results, provided important insights and helped with writing of the manuscript.
\section{Additional information}
The authors declare no competing financial interests. Supplementary information accompanies this paper. Correspondence and requests for materials should be addressed to W.G.v.d.W.

\end{document}